\documentstyle[12pt]{article}
\begin{document}
\begin{flushright}
WU-AP/56/95\\
qr-qc/9512046
\end{flushright}
\vskip 0.4cm
\baselineskip = 18pt
\begin{center}
{\large{\bf DINAMIC MONOPOLES AND SPACETIME STRUCTURE}}\footnote{
Presented at the poster session at 1st RESCUE International Symposium
THE COSMOLOGICAL CONSTANT AND THE EVOLUTION OF THE UNIVERSE, The
University of Tokyo, November 7-10, 1995, and based on Ref. 7.}

\vskip 0.3cm
{\sc Nobuyuki Sakai}\\
{\it Department of Physics, Waseda University, Shinjuku-ku, Tokyo 169,
Japan}
\end{center}

\vskip 0.2cm
\begin{abstract}

\baselineskip = 16pt
According to previous work on magnetic monopoles, static regular
solutions are nonexistent if the vacuum expectation value of the Higgs
field $\eta$ is larger than a critical value $\eta_{{\rm cr}}$, which is of
the order of the Planck  mass. In order to understand the properties of
monopoles for $\eta>\eta_{{\rm cr}}$, we investigate their dynamics
numerically and classify those dynamical solutions into three types as
follows. If $\eta$ is larger than another critical value $\eta_{{\rm inf}}
{}~(>\eta_{{\rm cr}})$, a monopole inflates and a wormhole structure appears
around it. In the case of $\eta_{{\rm cr}}<\eta<\eta_{{\rm inf}}$,
inflation does not occur and the dynamics depend on the ratio of the
Higgs self coupling constant $\lambda$ and the gauge coupling constant
$e^2$: if $\lambda/e^2\stackrel{<}{\sim}1$, a monopole just shrinks and
becomes a black hole; otherwise, a monopole approaches a stable configuration.
\end{abstract}

\vskip 0.2cm
\baselineskip = 16pt
\section{Introduction}

In recent years static and spherically symmetric solutions of the
Einstein-Yang-Mills-Higgs system have been intensively studied in the
literature (Breitenlohner et al. 1992; 1995, Lee et al. 1992, Ortiz
1992, Tachizawa et al. 1995). It was shown that particle-like regular
solutions exist only if the vacuum expectation value of the Higgs field
$\eta$ is less than a critical value $\eta_{{\rm cr}}$, which is of the
order of the Planck mass $m_{Pl}$. This result naturally gives rise to
the next question what is the fate of magnetic monopoles for
$\eta>\eta_{{\rm cr}}$. Because we cannot find an answer to the question
only by analysing static solutions, we investigate dynamic monopole
solutions in this paper. This subject is also related to the
``topological inflation" model, which was proposed by Linde (1994) and
Vilenkin (1994) independently.

\section{Results}

We consider the $SU(2)$ Einstein-Yang-Mills-Higgs system, which is
described by
$$
S=\int d^4 x \sqrt{-g} \left[\frac{m_{{\rm Pl}}^{~2}}{16\pi}{\cal R}
     -\frac14(F^a_{\mu\nu})^2
-\frac12(D_{\mu}\Phi^a)^2-{1\over
4}\lambda(\Phi^a\Phi^a-\eta^2)^2\right],
$$$$
F^a_{\mu\nu} \equiv \partial_{\mu}A^a_{\nu}-\partial_{\nu}A^a_{\mu}
  -e\epsilon^{abc}A^b_{\mu}A^c_{\nu},~~
D_{\mu}\Phi^a\equiv\nabla_{\mu}\Phi^a+e\epsilon^{abc}A^b_{\mu}\Phi^c,
$$
where $A^a_{\mu}$ and $F^a_{\mu\nu}$ are the $SU(2)$ Yang-Mills field
potential and its field strength, respectively. $\Phi^a$ is the real
triplet Higgs field. $\lambda$ and $e$ are the Higgs self coupling
constant and the gauge coupling constant, respectively. Assuming a
spherically symmetric spacetime and adopting the 't Hooft-Polyakov
ansatz for the matter fields, we solve dynamical equations numerically.

Contrary to the case of global monopoles, where all dynamic monopoles
undergo inflationary expansion without black-hole formation (Sakai et
al. 1996), we have found three types of solutions. Our results are
summarized in Fig. 1. A  triangle $(\triangle)$ denotes a stable
solution. A cross $(\times)$ denotes the case where a monopole shrinks.
A circle $(\circ)$ denotes the case where a monopole inflates and the
wormhole structure appears. A dotted line indicates the maximum values
of $\eta/m_{{\rm Pl}}$ versus $\lambda/e^2$, depicted approximately by
use of Fig.6 of Breitenlohner et al (1992).

We interpret the above results as follows. If $\eta$ is larger than
another critical value $\eta_{{\rm inf}} ~(>\eta_{{\rm cr}})$, which has
a little dependence on $\lambda/e^2$, a monopole inflates and a wormhole
structure appears around it. We sketch the spacetime structure for this
cases in Fig. 2; this is just a ``child universe" (Sato et al. 1982).
We should emphasize that a child universe can be generated without
fine-tuned initial conditions in this model, contrary to the case of a
trapped false vacuum bubble. In the case of $\eta_{{\rm
cr}}<\eta<\eta_{{\rm inf}}$, inflation does not occur and its dynamics
depend on $\lambda/e^2$: if $\lambda/e^2\stackrel{<}{\sim}1$, a monopole
just shrinks and becomes a black hole; otherwise, a monopole approaches
a stable configuration. The latter result indicates that there exist
stationary solutions even when no static solution exists, which is an
unexpected result.

\section{References}

\noindent
1. Breitenlohner, P., Forg$\grave{{\rm a}}$cs, P. and Maison, D., 1992,
{\it Nucl. Phys.} {\bf 383B}, 357.
\\\noindent
2. Breitenlohner, P., Forg$\grave{{\rm a}}$cs, P. and Maison, D., 1995,
{\it Nucl. Phys.} {\bf 442B}, 126.
\\\noindent
3. Lee, K., Nair, V.P., and Weinberg, E.J., 1992, {\it Phys.\ Rev.\ D}
{\bf 45}, 2751. \noindent
4. Linde, A.D., 1994, {\it Phys.\ Lett.\ B} {\bf 327}, 208.
\\\noindent
5. Ortiz, M.E., 1992, {\it Phys.\ Rev.\ D} {\bf 45}, R2586.
\\\noindent
6. Sakai, N., Shinkai, H., Tachizawa, T. and Maeda, K., 1996, {\it Phys.
\ Rev.\ D} {\bf 53}, to be published.
\\\noindent
7. Sakai, N., preprint, WU-AP/52/95, gr-qc/9512045.
\\\noindent
8. Sato, K., Kodama, H., Sasaki, M., and Maeda, K., 1982, {\it Phys.
\ Lett.\ B} {\bf108}, 103.
\\\noindent
9. Tachizawa, T., Maeda, K, and Torii, T., 1995, {\it Phys.\ Rev.\ D}
{\bf 51}, 4054.
\\\noindent
10. Vilenkin, A., 1994, {\it Phys.\ Rev.\ Lett.} {\bf 72}, 3137.

\end{document}